\documentstyle[aps]{revtex}
\title{REDUCTION OF THE N-COMPONENT SCALAR MODEL AT TWO LOOP LEVEL}
\author{Antal Jakov\'ac \\
{\em Department of Atomic Physics \\
  E\"otv\"os University\\
  Budapest, Hungary} \\}
\date{\today}
\newcommand{\tb}{&}
\newcommand{\nn}{\nonumber \\}
\def\P{\Phi_0}
\def\ch{{\cal H}}
\def\O{\Omega}
\def\loT#1{\log\left({#1\over T}\right)}
\def\lokT#1{\left(\log{#1\over T}\right)^2}
\def\lomu#1{\log\left({#1\over\mu}\right)}
\def\lokmu#1{\left(\log{#1\over\mu}\right)^2}

\def\pTint{T\sum\limits_{n\neq0}\int\!\!{d^3p\over(2\pi)^3}}
\def\ph{\varphi}
\def\N{\tilde N}
\def\tlam{\tilde\lambda}
\def\ep{\varepsilon}
\def\cirdef{
\multiput(30,0)(-0.0341,0.2588){30}{\makebox(0.4444,0.6667){\sevrm .}}
\multiput(28.98,7.76)(-0.0999,0.2412){30}{\makebox(0.4444,0.6667){\sevrm .}}
\multiput(25.98,15)(-0.1589,0.2071){30}{\makebox(0.4444,0.6667){\sevrm .}}
\multiput(21.21,21.21)(-0.2071,0.1589){30}{\makebox(0.4444,0.6667){\sevrm .}}
\multiput(15,25.98)(-0.2412,0.0999){30}{\makebox(0.4444,0.6667){\sevrm .}}
\multiput(7.76,28.98)(-0.2588,0.0341){30}{\makebox(0.4444,0.6667){\sevrm .}}
\multiput(0,30)(-0.2588,-0.0341){30}{\makebox(0.4444,0.6667){\sevrm .}}
\multiput(-7.76,28.98)(-0.2412,-0.0999){30}{\makebox(0.4444,0.6667){\sevrm .}}
\multiput(-15,25.98)(-0.2071,-0.1589){30}{\makebox(0.4444,0.6667){\sevrm .}}
\multiput(-21.21,21.21)(-0.1589,-0.2071){30}{\makebox(0.4444,0.6667){\sevrm .}}
\multiput(-25.98,15)(-0.0999,-0.2412){30}{\makebox(0.4444,0.6667){\sevrm .}}
\multiput(-28.98,7.76)(-0.0341,-0.2588){30}{\makebox(0.4444,0.6667){\sevrm .}}
\multiput(-30,0)(0.0341,-0.2588){30}{\makebox(0.4444,0.6667){\sevrm .}}
\multiput(-28.98,-7.76)(0.0999,-0.2412){30}{\makebox(0.4444,0.6667){\sevrm .}}
\multiput(-25.98,-15)(0.1589,-0.2071){30}{\makebox(0.4444,0.6667){\sevrm .}}
\multiput(-21.21,-21.21)(0.2071,-0.1589){30}{\makebox(0.4444,0.6667){\sevrm .}}
\multiput(-15,-25.98)(0.2412,-0.0999){30}{\makebox(0.4444,0.6667){\sevrm .}}
\multiput(-7.76,-28.98)(0.2588,-0.0341){30}{\makebox(0.4444,0.6667){\sevrm .}}
\multiput(0,-30)(0.2588,0.0341){30}{\makebox(0.4444,0.6667){\sevrm .}}
\multiput(7.76,-28.98)(0.2412,0.0999){30}{\makebox(0.4444,0.6667){\sevrm .}}
\multiput(15,-25.98)(0.2071,0.1589){30}{\makebox(0.4444,0.6667){\sevrm .}}
\multiput(21.21,-21.21)(0.1589,0.2071){30}{\makebox(0.4444,0.6667){\sevrm .}}
\multiput(25.98,-15)(0.0999,0.2412){30}{\makebox(0.4444,0.6667){\sevrm .}}
\multiput(28.98,-7.76)(0.0341,0.2588){30}{\makebox(0.4444,0.6667){\sevrm .}}
}
\def\dashcirdef{
\multiput(30,0)(-0.0341,0.2588){30}{\makebox(0.4444,0.6667){\sevrm .}}
\multiput(25.98,15)(-0.1589,0.2071){30}{\makebox(0.4444,0.6667){\sevrm .}}
\multiput(15,25.98)(-0.2412,0.0999){30}{\makebox(0.4444,0.6667){\sevrm .}}
\multiput(0,30)(-0.2588,-0.0341){30}{\makebox(0.4444,0.6667){\sevrm .}}
\multiput(-15,25.98)(-0.2071,-0.1589){30}{\makebox(0.4444,0.6667){\sevrm .}}
\multiput(-25.98,15)(-0.0999,-0.2412){30}{\makebox(0.4444,0.6667){\sevrm .}}
\multiput(-30,0)(0.0341,-0.2588){30}{\makebox(0.4444,0.6667){\sevrm .}}
\multiput(-25.98,-15)(0.1589,-0.2071){30}{\makebox(0.4444,0.6667){\sevrm .}}
\multiput(-15,-25.98)(0.2412,-0.0999){30}{\makebox(0.4444,0.6667){\sevrm .}}
\multiput(0,-30)(0.2588,0.0341){30}{\makebox(0.4444,0.6667){\sevrm .}}
\multiput(15,-25.98)(0.2071,0.1589){30}{\makebox(0.4444,0.6667){\sevrm .}}
\multiput(25.98,-15)(0.0999,0.2412){30}{\makebox(0.4444,0.6667){\sevrm .}}
}
\def\cira{\vcenter{\hbox{
\setlength{\unitlength}{0.5pt}
\begin{picture}(60,60)(-30,-30)
\cirdef
\multiput(25,-5)(0.5,0.5){20}{\makebox(0.4444,0.6667){\sevrm .}}
\multiput(25,5)(0.5,-0.5){20}{\makebox(0.4444,0.6667){\sevrm .}}
\end{picture}
}}}
\def\setsun{\vcenter{\hbox{
\setlength{\unitlength}{0.5pt}
\begin{picture}(60,60)(-30,-30)
\cirdef
\put(-30,0){\line(1,0){60}}
\end{picture}
}}}
\def\setsuna{\vcenter{\hbox{
\setlength{\unitlength}{0.5pt}
\begin{picture}(60,60)(-30,-30)
\cirdef
\put(-30,0){\line(1,0){60}}
\multiput(-60,0)(0.2611,0){30}{\makebox(0.4444,0.6667){\sevrm .}}
\multiput(-44.34,0)(0.2611,0){30}{\makebox(0.4444,0.6667){\sevrm .}}
\multiput(33.98,0)(0.2611,0){30}{\makebox(0.4444,0.6667){\sevrm .}}
\multiput(49.64,0)(0.2611,0){30}{\makebox(0.4444,0.6667){\sevrm .}}
\end{picture}
}}}
\def\setsunb{\vcenter{\hbox{
\setlength{\unitlength}{0.5pt}
\begin{picture}(60,60)(-30,-30)
\cirdef
\multiput(-60,0)(0.2611,0){30}{\makebox(0.4444,0.6667){\sevrm .}}
\multiput(-44.34,0)(0.2611,0){30}{\makebox(0.4444,0.6667){\sevrm .}}
\multiput(-28.67,0)(0.2611,0){30}{\makebox(0.4444,0.6667){\sevrm .}}
\multiput(-13.01,0)(0.2611,0){30}{\makebox(0.4444,0.6667){\sevrm .}}
\multiput(2.65,0)(0.2611,0){30}{\makebox(0.4444,0.6667){\sevrm .}}
\multiput(18.31,0)(0.2611,0){30}{\makebox(0.4444,0.6667){\sevrm .}}
\multiput(33.98,0)(0.2611,0){30}{\makebox(0.4444,0.6667){\sevrm .}}
\multiput(49.64,0)(0.2611,0){30}{\makebox(0.4444,0.6667){\sevrm .}}
\end{picture}
}}}
\def\setsunc{\vcenter{\hbox{
\setlength{\unitlength}{0.5pt}
\begin{picture}(60,60)(-30,-30)
\dashcirdef
\multiput(-60,0)(0.2611,0){30}{\makebox(0.4444,0.6667){\sevrm .}}
\multiput(-44.34,0)(0.2611,0){30}{\makebox(0.4444,0.6667){\sevrm .}}
\multiput(-28.67,0)(0.2611,0){30}{\makebox(0.4444,0.6667){\sevrm .}}
\multiput(-13.01,0)(0.2611,0){30}{\makebox(0.4444,0.6667){\sevrm .}}
\multiput(2.65,0)(0.2611,0){30}{\makebox(0.4444,0.6667){\sevrm .}}
\multiput(18.31,0)(0.2611,0){30}{\makebox(0.4444,0.6667){\sevrm .}}
\multiput(33.98,0)(0.2611,0){30}{\makebox(0.4444,0.6667){\sevrm .}}
\multiput(49.64,0)(0.2611,0){30}{\makebox(0.4444,0.6667){\sevrm .}}
\end{picture}
}}}
\def\fourptfnct{\vcenter{\hbox{
\setlength{\unitlength}{0.5pt}
\begin{picture}(60,60)(-30,-30)
\cirdef
\multiput(-30,0)(-0.2611,0.2621){30}{\makebox(0.4444,0.6667){\sevrm .}}
\multiput(-30,0)(-0.2611,-0.2621){30}{\makebox(0.4444,0.6667){\sevrm .}}
\multiput(-45.66,15.66)(-0.2611,0.2621){30}{\makebox(0.4444,0.6667){\sevrm .}}
\multiput(-45.66,-15.66)(-0.2611,-0.2621){30}{\makebox(0.4444,0.6667){\sevrm
.}}
\multiput(30,0)(0.2611,0.2621){30}{\makebox(0.4444,0.6667){\sevrm .}}
\multiput(30,0)(0.2611,-0.2621){30}{\makebox(0.4444,0.6667){\sevrm .}}
\multiput(45.66,15.66)(0.2611,0.2621){30}{\makebox(0.4444,0.6667){\sevrm .}}
\multiput(45.66,-15.66)(0.2611,-0.2621){30}{\makebox(0.4444,0.6667){\sevrm .}}
\end{picture}
}}}
\def\newvertex{\vcenter{\hbox{
\setlength{\unitlength}{0.5pt}
\begin{picture}(60,60)(-30,-30)
\put(-5,-2.5){\makebox(10,5){\vrule width 10pt height 5pt}}
\multiput(-10,5)(-0.2611,0.2621){30}{\makebox(0.4444,0.6667){\sevrm .}}
\multiput(-10,-5)(-0.2611,-0.2621){30}{\makebox(0.4444,0.6667){\sevrm .}}
\multiput(-25.66,20.66)(-0.2611,0.2621){30}{\makebox(0.4444,0.6667){\sevrm .}}
\multiput(-25.66,-20.66)(-0.2611,-0.2621){30}{\makebox(0.4444,0.6667){\sevrm
.}}
\multiput(10,5)(0.2611,0.2621){30}{\makebox(0.4444,0.6667){\sevrm .}}
\multiput(10,-5)(0.2611,-0.2621){30}{\makebox(0.4444,0.6667){\sevrm .}}
\multiput(25.66,20.66)(0.2611,0.2621){30}{\makebox(0.4444,0.6667){\sevrm .}}
\multiput(25.66,-20.66)(0.2611,-0.2621){30}{\makebox(0.4444,0.6667){\sevrm .}}
\end{picture}
}}}
\def\d{\partial}
\def\fel{{1\over2}}

\def\exv#1{\bigl<#1\bigr>}

\begin{document}
\draft
\maketitle

\begin{abstract}
Dimensional reduction of high temperature field theories improves IR
features of their perturbative treatment. A crucial question is, what
three-dimensional theory is representing the full system the most faithful
way. Careful investigation of the induced 3-dimensional counterterm structure
of the finite temperature 4D O(N) symmetric scalar theory at 2-loop level leads
to proposing the presence of non-local operators in the effective theory.
On scales beyond ${\cal O}(T^{-1})$, the scaling behavior of the couplings,
consequently, deviates from the usual three-dimensional scaling characteristic
for superrenormalizable theories.
\end{abstract}

\pacs{PACS numbers: 11.10.Wx, 64.60.-i, 95.30.Cq}

\section{Introduction}

Numerical investigation of the finite temperature electroweak phase
transition is unavoidable because in the symmetric phase no improved
perturbation method is known to work reliably. The most adequate approach
is to concentrate on the degrees of freedom left ''massless'', e.g. the static
Matsubara modes of the magnetic gauge and Higgs fluctuations \cite{KajBiel}.

The effective reduced theory usually is arrived at after simple 1-loop
integration over the nonstatic and the screened static modes
\cite{Farakos1,Jakovac}, with the restriction that only those terms are
retained in the effective action which are renormalizable in 4 dimensions.
These theories being superrenormalizable in 3 dimensions one can work out
their exact divergence structure and relate the physical, temperature dependent
mass to the bare parameters of the 3 dimensional theory \cite{Farakos2,Karsch}.
The general strategy behind this procedure is the matching of some
 important perturbatively computable quantities calculated both in the full
finite temperature theory and in the effective superrenormalisable
three-dimensional theory \cite{Braaten}.
The difficult and most interesting question is how accurate is this 3
dimensional superrenormalizable representation of the theory. On one hand one
should control higher dimensional operators with couplings inversely
proportional to some power of T \cite{Jakovac}. On the other hand on very
general grounds intrinsic non-local behaviour of range $T^{-1}$ is also
expected to appear.

The last point can be illustrated, for instance, on the example
 of the so-called "sunset"
diagram contributing to the propagation of a static mode.
The contributions are naturally divided into three groups:
\begin{equation}
\setsuna\qquad+\qquad\setsunc\qquad+\qquad\setsunb
\end{equation}
(the solid lines represent non-static, the dashed ones the static propagators).
The first class is accounted for when one integrates over the non-static modes
with two-loop accuracy. The second class is considered in the course of the
solution of the static effective model. The third ("mixed") class, however
would be left out from the two-step solution.

Our proposition is to incorporate the mixed part of the "sunset" contribution
into the effective model by defining a new kind of vertex, namely
\begin{equation}
\fourptfnct\qquad=\newvertex
\label{newvertex}\end{equation}
This vertex is clearly nonlocal, since the lines compressed into it have
a typical range ${\cal O}(T^{-1})$. Similar conclusion has been emphasized
recently also by  \cite{Mack}.

Explicit estimate of the strength of the
non-local corrections, to our knowledge will appear for the first time in
the present paper. In this first attempt we limit our calculations to the
N-component scalar model (for $N=4$ the Higgs-sector of the standard model)
in the hope, that the main lesson remains true for the full electroweak theory
\cite{Jakovac2}.

We did our computations using momentum cut-off regularization, that is
we regularized the propagator to be zero, when its momentum exceeds the
cut-off. The divergent counterterm structure obtained
for the effective model by  the
2-loop reduction and that of the 3 dimensional superrenormalisable theory
computed with cut-off regularization both display extra power divergences
which would be absent in dimensional regularization
\cite{Farakos1,Laine,Kripfganz}.
The need for extra pieces in the effective action becomes obvious
by the mismatch discovered just between those quantities. The correct choice
of the new terms will be undisputable if with their contributions
 the balance between three different type of power- and logarithmic
divergences of the effective theory and those induced by the reduction
will be reestablished.

Our investigation starts with a detailed presentation of the
counterterm structure of the 3 dimensional O(N) symmetric
model of the static modes
induced upon 2-loop integration over nonstatic degrees of freedom
(Section 2). This is then compared to the divergences of the effective
potential calculated from
the 3 dimensional local superrenormalizable O(N) model at the same
2-loop level (Section 3). A discrepancy between the results will be
discovered and the origin of the mismatch will be located in Section 4.
The extra divergencies needed for the consistency between the effective and
the original theory come from intrinsically non-local operators in 3
dimensions with a characteristic non-locality range $T^{-1}$,
as qualitatively explained above.
In the Conclusion (Section 5) we shall outline the matching
strategy for replacing the effective non-local theory by a local theory with
nonremovable cut-off $\alpha T^{-1}$ ($\alpha =1-5$). The interpretation of
the computer simulation of a corresponding lattice system will also be shortly
discussed. Also we shall touch upon the extension of the present
work to gauge theories.

\section{Reduction of the 4 dimensional O(N) model at two loop level}

The model is described by the following Hamiltonian:
\begin{equation}
\ch=\sum\limits_{i=1}^N\fel\left((\d\ph_i)^2+m^2\ph_i^2\right)
+{\lambda\over 4!}\left(\sum\limits_{i=1}^N\ph_i^2\right)^2+\ch_{ct},
\label{p4}\end{equation}
where $\ch_{ct}$ represents the countertems. First, we integrate out the
nonstatic Matsubara modes, and find the potential energy density for the
reduced
3 dimensional theory. For this we shift the fields $\ph$ by a constant
$\P$, what stands for the static part. Because of the $O(N)$ symmetry
there exists a coordinate system, where
$\P$ points along the N-th axis. The Hamiltonian breaks up into
$\ph$-independent, and quadratic pieces, and a higher power
$\ph$-dependent term:
\begin{equation}
\ch=\ch_0+\ch_2+\ch_I,
\end{equation}
where
\begin{eqnarray}
\tb\ch_0\tb=\fel(m^2+\delta m^2)\P^2+{1\over 4!}(\lambda+\delta\lambda)\P^4,\nn
\tb\ch_2\tb=\sum\limits_{i=1}^{N-1}\fel\ph_i\left(-\d^2+m^2+{\lambda\over6}
\P^2\right)\ph_i+\fel\ph_N\left(-\d^2+m^2+{\lambda\over2}\P^2\right)\ph_N,\nn
\tb\ch_I\tb=\sum\limits_{i=1}^{N-1}\fel\ph_i\left(-(Z-1)\d^2+\delta m^2+{\delta
\lambda\over6}\P^2\right)\ph_i+\fel\ph_N\left(-(Z-1)\d^2+\delta m^2
+{\delta\lambda\over2}\P^2\right)\ph_N\nn
\tb\tb+{\lambda+\delta\lambda\over4!}\left(\left(
\sum\limits_{i=1}^N\ph_i^2\right)^2+2\ph_N^2\sum\limits_{i=1}^{N-1}\ph_i^2
+4\P\ph_N\sum\limits_{i=1}^{N-1}\ph_i^2+\ph_N^4+4\P\ph_N^3\right).
\end{eqnarray}
We compute the potential energy density of the effective theory $f$
in the following way:
\begin{equation}
fV=H_0+\fel Tr\log\Delta-\exv{e^{-H_I}-1}_c,
\end{equation}
where the index $c$ refers to connected graphs, V is the volume of the system.

For two loop results the interaction part of the exponential must be expanded
to ${\cal O}(\lambda^2)$, and only $\P$-dependent terms should be retained:
\begin{eqnarray}
\tb-\exv{e^{-H_I}-1}_c=\tb\sum\limits_{i=1}^N\fel\left(\delta m^2+{\delta
\lambda\over6}\P^2\right)\exv{\ph_i^2}_c+\fel\left(\delta m^2+{\delta
\lambda\over2}\P^2\right)\exv{\ph_N^2}_c\nn
\tb\tb+{\lambda\over24}\left[\exv{\left(\sum\limits_{i=1}^{N-1}\ph_i^2
\right)^2}_c+2\exv{\ph_N^2\sum\limits_{i=1}^{N-1}\ph_i^2}_c+\exv{\ph_N^4}_c
\right]\nn
\tb\tb-{\lambda^2\over72}\P^2\exv{\ph_N(x)\sum\limits_{i=1}^N\ph_i^2(x)
\ph_N(y)\sum\limits_{i=1}^N\ph_i^2(y)}_c
\end{eqnarray}

This result can be rewritten in terms of Feynman-graphs.
We introduce the following notations
\begin{eqnarray}
\tb\tb I(m):={1\over V}Tr\log\Delta,\nn
\tb\tb K(m_1,m_2,m_3):=\setsun,
\label{graphs}\end{eqnarray}
where $m_1,\,m_2,\,m_3$ are the masses of the propagators on the three internal
lines of the diagram. It is clear, that
\begin{equation}
\cira={\d I(m)\over\d m^2}=:I'(m).
\end{equation}
We introduce the notations
\begin{eqnarray}
\tb\tb m_G^2:=m^2+{\lambda\over6}\P^2,\nn
\tb\tb m_H^2:=m^2+{\lambda\over2}\P^2,
\end{eqnarray}
and get the following result:
\begin{eqnarray}
\tb f=\tb \fel(m^2+\delta m^2)\P^2+{\lambda+\delta\lambda\over24}\P^4
+\fel(N-1)I(m_G)+\fel I(m_H)+\fel(\delta m^2+{\delta\lambda\over6})
(N-1)I'(m_G)\nn
\tb\tb+\fel(\delta m^2+{\delta\lambda\over2})I'(m_H)
+{\lambda\over24}\left[(N^2-1) I'(m_G)^2 + 2(N-1)I'(m_G)I'(m_H)
+3 I'(m_H)^2\right]\nn
\tb\tb-{\lambda^2\over36}\P^2\left[3 K(m_H,m_H,m_H)+(N-1)K(m_H,m_G,m_G)\right]
\label{twoloop}\end{eqnarray}

The functions $I$ and $K$ can be expanded with respect to the masses, if
no IR divergencies arise. This happens in our case, where the IR sensitive part
is substracted because no static mode is allowed to propagate
 on the internal lines.
Therefore, one is allowed to expand $I$ and $K$ in powers of  $m^2/T^2$. If we
don't want to keep the operators suppressed by some inverse power of $T$,
we can truncate the expansion after the
first few terms (high temperature expansion):
\begin{eqnarray}
\tb\tb I(m)=I_0+I_1 m^2+I_2 m^4+\dots,\nn
\tb\tb I'(m)=I_1+2I_2 m^2+3I_3 m^4+\dots,\nn
\tb\tb K(m_1,m_2,m_3)=K_0+K_1 {m_1^2+m_2^2+m_3^2\over3}+\dots
\label{eq:IKdef}\end{eqnarray}
(the function $K$ is symmetric in the three $m$'s). Relying on the divergence
structure of the zero temperature model, and
on dimensional analysis, the expected
form of the coefficients in (\ref{eq:IKdef}) is:
\begin{eqnarray}
I_1\tb=\tb I_1^2 \Lambda^2+I_1^1 T\Lambda+I_1^0 T^2,\nn
I_2\tb=\tb I_2^{log}\loT{\Lambda}+I_2^0,\nn
I_3\tb=\tb {I_3\over T^2},\nn
K_0\tb=\tb K_0^2\Lambda^2+K_0^1 T\Lambda+K_0^{log}T^2\loT{\Lambda}+K_0^0
T^2,\nn
K_1\tb=\tb K_1^{2log}\lokT{\Lambda}+K_1^{log}\loT{\Lambda}+K_1^0.
\label{eq:divstr}\end{eqnarray}

For future convenience let us introduce $\N=(N+2)/3$. At one loop level,
if we choose the counterterms (in general renormalization scheme) as
\begin{eqnarray}
\tb\tb \delta m_1^2=\Delta m_1^2-\N\lambda\left(\fel I_1^2\Lambda^2+
m^2 I_2^{log} \lomu{\Lambda}\right),\nn
\tb\tb \delta\lambda_1=\Delta\lambda_1-(\N+2)\lambda^2 I_2^{log}\lomu{\Lambda},
\label{oneloopct}\end{eqnarray}
where the terms $\Delta$ are finite (and specify the renormalisation scheme),
we get for the effective potential:
\begin{eqnarray}
f\tb=\tb\fel\P^2\left(m^2+\Delta m_1^2+\N\lambda\left(\fel
I_1^0 T^2+I_2^0 m^2\right)+\N\lambda m^2 I_2^{log}\loT{\mu}
+{\N\over2}\lambda I_1^1 T\Lambda\right)\nn
\tb\tb+{1\over24}\P^4\left(\lambda+\Delta\lambda_1+(\N+2)\lambda^2 I_2^0+
(\N+2)\lambda^2 I_2^{log}\loT{\mu}\right)+\dots
\end{eqnarray}

At two loop level, with the same procedure we choose the counterterms as
\begin{eqnarray}
\tb\delta m_2^2\tb=\Delta m_2^2-\N\Lambda^2\left(\fel\Delta\lambda_1 I_1^2
-{1\over6}K_0^2\lambda^2-{\N+2\over2}I_1^2I_2^{log}\lambda^2\lomu{\Lambda}
\right)\nn
\tb\tb+\N\left((\N+2)(I_2^{log})^2+{1\over6}K_1^{2log}\right)\lambda^2m^2
\lokmu{\Lambda}\nn
\tb\tb+\N\left(2(I_2^{log})^2+{1\over3}K_1^{2log}\right)\lambda^2m^2
\lomu{\Lambda}\loT{\mu}\nn
\tb\tb-\N\left(\Delta m_1^2I_2^{log}\lambda+\Delta\lambda_1I_2^
{log}m^2 -2I_2^{log}I_2^0\lambda^2m^2-{1\over6}K_1^{log}\lambda^2m^2\right)
\lomu{\Lambda},\nn
\tb\delta\lambda_2\tb=\Delta\lambda_2
+\left({20+22\N+3\N^2\over3}(I_2^{log})^2+{5\N+4\over9}K_1^{2log}\right)
\lambda^3 \lokmu{\Lambda}\nn
\tb\tb+2{5\N+4\over9}\left(6(I_2^{log})^2+K_1^{2log}\right)
\lambda^3\lomu{\Lambda}\loT{\mu}\nn
\tb\tb-\left(2(\N+2)\Delta\lambda_1I_2^{log}\lambda-4{5\N+4\over3}I_2^{log}
I_2^0\lambda^3-{5\N+4\over9}K_1^{log}\lambda^3\right)\lomu{\Lambda}.
\end{eqnarray}
Using also the expression of the one loop counterterms (\ref{oneloopct}) we
obtain the 2-loop result valid in any general renormalization scheme.
Some important systematics can be discovered in this expression (see the
comments below eq.(\ref{twoloopf})),
 if one introduces the following shorthand notation:
\begin{equation}
\tlam=\lambda+\Delta\lambda_1+\N I_2^0\lambda^2+(\N+2)I_2^{log}\lambda^2
\loT{\mu}.
\end{equation}
Then the two-loop effective potential, renormalised from the point of view
of four-dimensional ultraviolet behavior can be written with
${\cal O}(\lambda^3)$ accuracy as
\begin{eqnarray}
f\tb=\tb\fel\P^2\biggl[(m^2+\Delta m_2^2)\left(1+\N\tlam\left(I_2^0+I_2^{log}
\loT{\mu}\right)\right)+\Delta m_1^2+m^2\N\tlam^2\biggl({3\N\over2}I_1^0I_3
-{1\over6}K_1^0\nn
\tb\tb-\left(2I_2^0I_2^{log}+{1\over6}K_1^{log}\right)\loT{\mu}
-\left(2(I_2^{log})^2+{1\over6}K_1^{2log}\right)\lokT{\mu}\biggr)\nn
\tb\tb +\N T^2\tlam\left(\fel I_1^0-{1\over6}K_0^0\tlam-\N\tlam(I_1^0I_2^{log}
+{1\over6}K_0^{log})\loT{\Lambda}\right)\nn
\tb\tb +\N\tlam\Lambda T\left(\fel I_1^1 -{1\over6}K_0^1\tlam-
\tlam I_1^1 I_2^{log}\loT{\Lambda}\right)\biggr]\nn
\tb+\tb{1\over24}\P^4\biggl[\tlam+2I_2^0\tlam^2+\Delta\lambda_2
+2\tlam\Delta\lambda_1\left(\N I_2^0+(\N+2)I_2^{log}\loT{\mu}\right)
+{3\N^2-10\N+4\over3}(I_2^0)^2\tlam^3\nn
\tb\tb+{3\over2}\N(\N+2)I_1^0I_3\tlam^3
-{5\N+4\over9}K_1^0\tlam^3+\tlam^3\loT{\mu}\biggl(
2{3\N^2-4\N-8\over3}I_2^{log}I_2^0-{5\N+4\over9}K_1^{log}\biggr)\nn
\tb\tb+\tlam^3\lokT{\mu}\biggl({3\N^2+2\N+4\over3}(I_2^{log})^2
-{5\N+4\over9}K_1^{2log}\biggr)\biggr]
\label{twoloopred}\end{eqnarray}

If we choose $\mu=T$, to avoid large logarithms, and a special scheme, eg.
the following (Weinberg-type) one
\footnote{The correct way would be to choose $\mu$ to be equal to
the scale where the parameters $\lambda$ and $m$ are fixed (eg. the mass
of the Higgs boson in the present case). The large logarithms can be summed
with the help of RG. This would give the above result with logarithmically
T-dependent $\lambda$ and $m$.}
\begin{eqnarray}
\tb\tb{\d^2f\over\d\P^2}\Biggr|_{\mu=T=0,\P=0}\!\!\!\!\!\!\!\!\!:=m^2,\nn
\tb\tb{\d^4f\over\d\P^4}\Biggr|_{\mu=T=0,\P=0}\!\!\!\!\!\!\!\!\!:=\lambda,
\end{eqnarray}
we can obtain a considerably simpler result. At one loop level
\begin{equation}
f=\fel\P^2\left(m^2+\fel\N\lambda I_1^0 T^2+\fel\N\lambda I_1^1 T\Lambda\right)
+{1\over24}\lambda\P^4+\dots
\end{equation}
which corresponds to the following choice of the finite parts:
\begin{eqnarray}
\tb\tb\Delta m_1^2=-\N\lambda I_2^0 m^2,\nn
\tb\tb\Delta\lambda_1=-(\N+2)\lambda^2 I_2^0.
\end{eqnarray}
One might notice, that at 1-loop the coefficients of $T^2$ and of $\Lambda T$
do not depend on the details, how the regularising cut-off is imposed.

At two loop level (without explicitly giving the complicated expressions
for $\Delta m_2^2$ and $\Delta\lambda_2$)
\begin{eqnarray}
f\tb=\tb\fel\P^2\biggl[m^2
+\N T^2\left(\fel\lambda I_1^0-\lambda^2\left(I_1^0 I_2^0+{1\over6}K_0^0
+\left(I_1^0 I_2^{log}+{1\over6}K_0^{log}\right)\loT{\Lambda}\right)\right)\nn
\tb\tb+\N\Lambda T\left(\fel\lambda I_1^1-\lambda^2\left(I_1^1I_2^0
+{1\over6}K_0^1+I_1^1I_2^{log}\loT{\Lambda}\right)\right)\biggr]
+{1\over24}\lambda\P^4.
\label{twoloopf}
\end{eqnarray}
It can be seen from formula (\ref{twoloopred}), that the coefficients of the
terms $\Phi^2T^2$ and $\Phi^2\Lambda T$ do not depend on the renormalisation
scheme and scale ($\mu)$. That means, that essentially this form would have
arrived also at any other renormalisation scheme. On the other hand
the coefficient of $\Phi^2 T^2$ and $\Phi^2\Lambda T$ at ${\cal O}(\lambda^2)$
depend on the specific implementation of the cut-off (regularisation
dependence), through the coefficients $K_0^1$ and $K_0^0$. Therefore
for computing physical quantities it is necessary to apply a common unique
regularization procedure, otherwise the 3D linear divergence would not be
cancelled and also some ${\cal O}(\lambda^2)$ finite contributions
would be  unreliable.
The coefficients of other induced divergencies of the type
 $T^2\log(\Lambda/T),~\Lambda T \log(\Lambda/T)$ do not depend on the specific
implementation of the cut-off in multiloop integrals.

The only remaining task is the computation of the constants defined in
(\ref{eq:divstr}). The $I_i$ quantities come from the tadpole graph:
\begin{equation}
I_1+2I_2M^2+3I_3M^4+\dots=\pTint{1\over \omega_n^2+p^2+M^2}.
\end{equation}
The $K_i$ constants are determined by the setting sun diagram:
\begin{equation}
K_0+K_1M^2+K_2M^4+\dots=T^3\sum\limits_{n,l,m\neq0}\int\!\!{d^3pd^3qd^3k
\over(2\pi)^9}\,{\beta(2\pi)^3\delta_{n+m+l,0}\delta^{(3)}(p+q+k)
\over(\omega_n^2+E_p^2)(\omega_m^2+E_q^2)(\omega_l^2+E_k^2)}.
\label{eq:setsun}\end{equation}
where $E_p^2=p^2+M^2$.
The result of the integrations (see Appendix A and B) are:
\begin{eqnarray}
I_1\tb=\tb{\Lambda^2\over8\pi^2}-{\Lambda T\over2\pi^2}+{T^2\over12},\nn
I_2\tb=\tb-{1\over16\pi^2}\loT{\Lambda}+{1\over16\pi^2}(1+\log(2\pi)-
\gamma_E),\nn
I_3\tb=\tb{\zeta(3)\over192\pi^4T^2},\nn
K_0\tb=\tb0.0001041333\,\Lambda^2-0.0029850437\,\Lambda T+
{5\over32\pi^2}T^2\loT{\Lambda}-0.0152887686\,T^2,\nn
K_1\tb=\tb-{3\over128\pi^4}\lokT{\Lambda}
+0.001087871\,\loT{\Lambda}+{\rm const.}
\label{constants}\end{eqnarray}
The coefficients written in decimal form are results of numerical integrations,
other coefficients were calculated analytically. The former depend on the
special implementation of the cut-off procedure the latter do not.

With these values we obtain the following result for
the 3 dimensional effective
potential in Weinberg-type renormalization scheme:
\begin{eqnarray}
f\tb=\tb\fel\P^2\biggl[m^2+\N T^2\biggl({\lambda\over24}
-0.0013551443\,\lambda^2-{\lambda^2\over48\pi^2}\loT{\Lambda}\biggr)\nn
\tb\tb+\N\Lambda
T\left(-{\lambda\over4\pi^2}+0.0012227544\,\lambda^2-{\lambda^2
\over32\pi^2}\loT{\Lambda}\right)\biggr]+{\lambda\over24}\P^4.
\label{numred}\end{eqnarray}
The one loop result is well known \cite{Farakos1,Jakovac}, the two loop result
is new. From this expression one can read off the induced 3 dimensional mass
counterterm:
\begin{equation}
\delta m^2_{ind}=-\N\left({\lambda\over4\pi^2}(1-0.048277409\,\lambda)\Lambda T
+{\lambda^2\over32\pi^2}\Lambda T\loT{\Lambda}
+{\lambda^2\over48\pi^2}T^2\loT{\Lambda}\right)
\label{pred}\end{equation}
No new divergence related to $\lambda$ appears, supporting the
superrenormalizable nature of the 3 dimensional theory. We have to compare
this induced counterterm to the counterterm required for the
3 dimensional theory of the same form as (\ref{p4}) expected on
the basis of superrenormalizability. The eventual cancellation is a
precondition
for the consistent representation of the 4D finite T theory by
this simplest minded 3D form. But as we shall see this hope doesn't come true.

\section{2-loop effective potential of the 3D Higgs-model}

The result (\ref{numred}) of the previous section  suggests a 3D effective
representation
of the original theory which could be of the form (\ref{p4}).
This is again a local scalar O(N) model, just now with
temperature dependent mass parameter. The renormalization of the theory,
however, on the 4D level has been accomplished already.
The counterterms of the "embedded" 3D theory
were generated by the reduction.
A crucial check of consistency on the suggested model
 is the cancellation of its divergencies
against the induced counterterms.

In this section we investigate the effective potential of the {\it three}
 dimensional
theory defined by the action (\ref{p4}) at two loop level, when the bare mass
is temperature dependent, as described by (\ref{numred}). The results are known
for even more complex theories \cite{Farakos1}, so this section will be
just a short summary of well-known facts.

In three dimensions the theory is superrenormalizable, we expect divergent
counterterm only to the mass parameter, and only up to two loop level.
Otherwise the graphs are the same as in the 4 dimensional case
(see (\ref{graphs})), but the integrals are 3 dimensional. The result of the
integration is (cf. (\ref{twoloop})):
\begin{eqnarray}
\tb f=\tb \fel(m^2+\delta m^2)\P^2+{\lambda\over24}\P^4
+\fel(N-1)I(m_G)+\fel I(m_H)+\fel \delta m^2 (N-1)I'(m_G)\nn
\tb\tb+\fel \delta m^2 I'(m_H)
+{\lambda\over24}\left[(N^2-1) I'(m_G)^2 + 2(N-1)I'(m_G)I'(m_H)
+3 I'(m_H)^2\right]\nn
\tb\tb-{\lambda^2\over36}\P^2\left[3 K(m_H,m_H,m_H)+(N-1)K(m_H,m_G,m_G)\right].
\label{thrdim}\end{eqnarray}
On dimensional reasons, and using the expected form of the
divergencies we get for the relevant parts of the functions $I$ and $K$
(Don't forget that we deal with three-dimensional theory!):
\begin{eqnarray}
\tb\tb I(m)=2J_1\Lambda m^2+2J_0 m^3,\nn
\tb\tb K(m,m,M)=L_{log}\log{\Lambda^2\over\mu^2}+L_0-2L_{log}\log
{2m+M\over\mu}.
\end{eqnarray}
Let us choose the counterterm
\begin{equation}
\delta m_1^2=\Delta m_1^2-{1\over3}(N+2)J_1\lambda\Lambda,
\label{thr1ct}\end{equation}
what leads for the one loop part of (\ref{thrdim}) to the result:
\begin{equation}
v^{eff}_{1ren}=\fel(m^2+\Delta m_1^2)\P^2+{\lambda\over24}\P^4
+J_0\left(m^2+{\lambda\over2}\P^2\right)^{3/2}
+(N-1)J_0\left(m^2+{\lambda\over6}\P^2\right)^{3/2}.
\end{equation}
Using (\ref{thr1ct}) combined with the two loop counterterm
\begin{equation}
\delta m_2^2=\Delta m_2^2+{\lambda^2\over18}(N+2)L_{log}
\log{\Lambda^2\over\mu^2},
\label{thr2ct}\end{equation}
the two-loop effective potential (\ref{thrdim}) becomes:
\begin{eqnarray}
v_{2ren}^{eff}\tb=\tb\fel\left(m^2+\Delta m_1^2+\Delta m_2^2
-{\lambda^2\over18}(N+2)L_0+{\lambda^2\over8}(N^2+8)J_0^2\right)\P^2
+{\lambda\over24}\P^4\nn
\tb\tb+J_0\left(m^2+{\lambda\over2}\P^2\right)^{3/2}
+(N-1)J_0\left(m^2+{\lambda\over6}\P^2\right)^{3/2}
+{3\over2}\Delta m_1^2J_0\left(m^2+{\lambda\over2}\P^2\right)^{1/2}\nn
\tb\tb+{3\over2}(N-1)\Delta m_1^2J_0\left(m^2+{\lambda\over6}\P^2\right)^{1/2}
+{3\over4}(N-1)\lambda J_0^2\left(m^2+{\lambda\over2}\P^2\right)^{1/2}
\left(m^2+{\lambda\over6}\P^2\right)^{1/2}\nn
\tb\tb+{\lambda^2\over18}\P^2L_{log}\left[3\log{3m_H\over\mu}
+(N-1)\log{2m_G+m_H\over\mu}\right].
\end{eqnarray}
We can choose, in particular, the scheme
\begin{eqnarray}
\tb\Delta m_1^2\tb=0,\nn
\tb\Delta m_2^2\tb=\lambda^2\left({N+2\over18}L_0-{N^2+8\over8}J_0^2\right),
\end{eqnarray}
which gives for the finite potential
\begin{eqnarray}
\tb v_{2MS}^{eff}\tb=\fel m^2\P^2+{\lambda\over24}\P^4
+J_0\left(m^2+{\lambda\over2}\P^2\right)^{3/2}
+(N-1)J_0\left(m^2+{\lambda\over6}\P^2\right)^{3/2}\nn
\tb\tb+{3\over4}(N-1)\lambda J_0^2\left(m^2+{\lambda\over2}\P^2\right)^{1/2}
\left(m^2+{\lambda\over6}\P^2\right)^{1/2}\nn
\tb\tb+{\lambda^2\over18}\P^2L_{log}\left[3\log{3m_H\over\mu}
+(N-1)\log{2m_G+m_H\over\mu}\right].
\end{eqnarray}

Explicitly performing the 3D tadpole and setting sun calculations (see
Appendix A), the values of the constants are:
\begin{eqnarray}
\tb\tb J_1={1\over4\pi^2},\nn
\tb\tb J_0=-{1\over12\pi},\nn
\tb\tb L_{log}={1\over32\pi^2},\nn
\tb\tb L_0=6.70322\cdot10^{-3}.
\end{eqnarray}
With these values
\begin{eqnarray}
\tb v_{2MS}^{eff}\tb=\fel m^2\P^2+{\lambda\over24}\P^4-{1\over12\pi}\left[
\left(m^2+{\lambda\over2}\P^2\right)^{3/2}\!\!
+(N-1)\left(m^2+{\lambda\over6}\P^2\right)^{3/2}\right]\nn
\tb\tb+{N-1\over192\pi^2}\lambda\left(m^2+{\lambda\over2}\P^2\right)^{1/2}
\left(m^2+{\lambda\over6}\P^2\right)^{1/2}\nn
\tb\tb+{\lambda^2\over576\pi^2}\P^2\left[3\log{3m_H\over\mu}
+(N-1)\log{2m_G+m_H\over\mu}\right].
\end{eqnarray}

The divergence structure of the 3D theory can be read off from
(\ref{thr1ct}) and (\ref{thr2ct}):
\begin{equation}
v_{div}={N+2\over3}\left[{\lambda\over4\pi^2}\Lambda
-{\lambda^2\over192\pi^2}\log{\Lambda^2\over\mu^2}\right]\Phi_0^2.
\label{divloc}\end{equation}
This coincides with the result of \cite{Farakos1}, if we change
$\lambda\to\lambda/6$, which corresponds to the choice of the coefficient of
the scalar selfinteraction to be $\lambda/4$.

The mismatch between the induced  counterterms  (\ref{pred}) and the
three-dimensional  divergencies (\ref{divloc}) is by now quite obvious.
There are
some divergences present in (\ref{pred}), which don't appear in (\ref{divloc})
(eg. $\sim \Lambda T\log(\Lambda/T)$), but even the operators present in both
cases have {\sl different} coefficients (eg. $\sim T^2log(\Lambda/T)$).
This means, that in a regularisation scheme, where power-like divergences
are simply subtracted (see \cite{Braaten}), one still could recognize the need
for supplementary terms in the action of the effective theory.
The divergence structure of the reduced theory is richer anyhow than we could
expect from the simplest (superrenormalisable)
representation of the theory, eg. the scale
dependence (beta-function) of the faithfully reduced theory is different.
For the resolution of this puzzle, and for reconstructing the correct
scale dependence we have to introduce some new operators into the action of the
effective model. This will be proposed in the following section.

\section{Filling the gap: the one loop four point function}

To cure the disease of the naively constructed 3D model
detected in the previous section we have to add some new
operators to the basic (\ref{p4}) action. The arguments in the Introduction
suggest to investigate the four-point function (\ref{newvertex}) and substitute
it by a new vertex. Since, already all momentum-independent
operators with couplings proportional to non-negative powers of T are taken
into account, the new operators have to be momentum-dependent.
 From the point of view of the full theory, as we argued in the Introduction,
the new term should represent the diagrams with mixed static and non-static
internal lines.
Such operators
can be generated only radiatively, namely in the process of the integration
over the nonstatic modes at 1-loop. In a subsequent 3D 1-loop computation of
its contribution to the effective potential, 2-loop divergencies
yet missing in the comparison of the induced counterterms with the
divergencies of the 3D local effective theory should emerge.

We are going to evaluate the 1-loop correction to the 4-point function
with static external lines of non-zero spatial momentum. When contracting
two of its external lines as part of the solution of the effective model
one recognizes the "mixed" static-nonstatic sunset contribution to the
static 2-point function.

\hbox to \hsize{
\vbox{\hsize 5truecm
\setlength{\unitlength}{0.0125in}
\begin{picture}(0,90)(100,680)
\put(160,720){\circle{80}}
\put(137,720){\line(-1,-1){ 20}}
\put(137,720){\line(-1, 1){ 20}}
\put(183,720){\line( 1,-1){ 20}}
\put(183,720){\line( 1, 1){ 20}}
\put(122,710){\line( 1, 0){  5}}
\put(127,705){\line( 0, 1){  5}}
\put(122,730){\line( 1, 0){  5}}
\put(127,735){\line( 0,-1){  5}}
\put(193,730){\line( 0, 1){  5}}
\put(193,730){\line( 1, 0){  5}}
\put(193,710){\line( 1, 0){  5}}
\put(193,710){\line( 0,-1){  5}}
\multiput(160,698)(0.5,0.5){7}{\makebox(0.4444,0.6667){\sevrm .}}
\multiput(160,698)(0.5,-0.5){7}{\makebox(0.4444,0.6667){\sevrm .}}
\multiput(160,742)(-0.5,0.5){7}{\makebox(0.4444,0.6667){\sevrm .}}
\multiput(160,742)(-0.5,-0.5){7}{\makebox(0.4444,0.6667){\sevrm .}}
\put(95,730){\makebox(0,0)[lb]{\raisebox{0pt}[0pt][0pt]
{$\scriptstyle(p_1,0)$}}}
\put(95,705){\makebox(0,0)[lb]{\raisebox{0pt}[0pt][0pt]
{$\scriptstyle(p_2,0)$}}}
\put(203,730){\makebox(0,0)[lb]{\raisebox{0pt}[0pt][0pt]
{$\scriptstyle(p_3,0)$}}}
\put(203,705){\makebox(0,0)[lb]{\raisebox{0pt}[0pt][0pt]
{$\scriptstyle(p_4,0)$}}}
\put(150,687){\makebox(0,0)[lb]{\raisebox{0pt}[0pt][0pt]
{$\scriptstyle(p,n)$}}}
\put(135,750){\makebox(0,0)[lb]{\raisebox{0pt}[0pt][0pt]
{$\scriptstyle(p+p_1+p_2,n)$}}}
\end{picture}
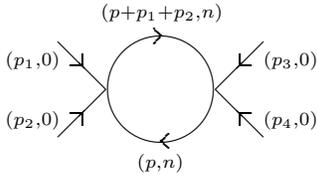
\centerline{\bf Fig. 1}
\vskip5pt
}
\raise0.3truecm\vbox{\hsize 9.7truecm
The relevant integral ($k=p_1+p_2$) is seen to be:
\begin{equation}
I_\Lambda(k)=\int_p^{\prime\Lambda}{1\over p^2+(2\pi nT)^2}{1\over(p+k)^2
+(2\pi nT)^2},
\end{equation}
}
\hfill}

\noindent where the primed integral is an abbreviation for the cut-off
integration with respect to $p$
\begin{equation}
\int_p^{\prime\Lambda}=T\sum\limits_{n\neq0}\int\!\!{d^3p\over(2\pi)^3}.
\end{equation}
With Feynman-parametrization one rewrites it as:
\begin{equation}
I_\Lambda(k)=\int\limits_0^1\!dy\,\int_p^{\prime\Lambda}
{1\over(p^2+(2\pi nT)^2+k^2y(1-y))^2},
\end{equation}

If the external lines are labelled by i, j, k ,l respectively, the tensorial
structure of the graph in Fig. 1. is proportional to
\begin{equation}
(N+4)\delta_{ij}\delta_{kl}+2(\delta_{ik}\delta_{jl}+\delta_{il}\delta_{jk}).
\label{tensstr}\end{equation}
Having this tensorial decomposition
 it is enough to work with homogeneous background as in the
previous sections.

The result of the integration is known for $k^2=0$:
\begin{equation}
I_\Lambda(k=0)={D_0-1\over8\pi^2}={1\over8\pi^2}\left(\log{\Lambda\over T}
-\log2\pi+\gamma_E-1\right).
\label{smallk}\end{equation}

In the other limit, $k^2\gg T^2$, one can look for an expansion in positive
powers of T (there is no other scale in the integral). Using the
Euler-Maclaurin resummation formula:
\begin{equation}
\sum\limits_{i=0}^nf(i)=\int\limits_0^n\!dx\,f(x)+\fel[f(0)+f(n)]+
\sum\limits_{k=1}^\infty{B_{2k}\over(2k)!}[f^{(2k-1)}(n)-f^{(2k-1)}(0)]
\end{equation}
(where $B_{2k}$'s are the Bernoulli-numbers) we get the following result:
\begin{equation}
I_\Lambda(k\gg T)={1\over8\pi^2}\log{2\Lambda\over k}-{1\over8\pi^2}
F\left({k\over\Lambda}\right)-{1\over8}{T\over k}
-{1\over16\pi^2}\sum\limits_{j=1}^\infty{(-1)^j(j-1)!B_{2j}\over(2j)!}
\left({32\pi^2T^2\over k^2}\right)^j,
\label{largek}\end{equation}
where
\begin{equation}
F(x)=\left({2\over x}-1\right)\ln\left(1-{x\over2}\right)+1.
\end{equation}

Terms proprotional $1/k^n$ ($n\ge2$) give constant contribution in three
dimensions -- these are the finite corrections to the four point function.
The first three terms, however, might lead to divergencies when in a subsequent
1-loop calculation two of the external lines of Fig.1 are contracted. If we
want to separate only the divergent contribution to the effective potential
upon contraction of two legs of Fig.1, we can interpolate between the relevant
pieces of (\ref{smallk}) and (\ref{largek}) with help of the following
expression:
\begin{equation}
I_\Lambda(k)\approx{1\over8\pi^2}\left[\log{\Lambda\over CT}-\O(k)
\right],
\label{intpol}\end{equation}
where $C=2\pi e^{1-\gamma_E}$ and
\begin{equation}
\O(k)={\pi^2Tk\over T^2+k^2}+\fel\log\left(1+{k^2\over4C^2T^2}\right)
+F\left({k\over\Lambda}\right).
\label{omega}\end{equation}

In section III,
we have obtained for the one loop level momentum-independent radiative
correction (at  $k=0$) to the potential energy density from Fig.1:
\begin{equation}
\Delta V_{3D}=-{\N+2\over3}{\lambda^2\over128\pi^2}\left[\log{\Lambda\over T}
-\log2\pi+\gamma_E-1\right]\P^4.
\end{equation}
This will now be modified for x-dependent fields $\P$ by adding the operators:
\begin{equation}
\Delta{\cal L}_{3D}={\N+2\over3}{\lambda^2\over128\pi^2}\P^2\O(i\d)\P^2.
\label{opadd}\end{equation}
The Fourier-transform of (\ref{opadd}) promptly reproduces the k-dependent part
of (\ref{intpol}).
For the modification of the original theory (\ref{p4}) we have to project back
(\ref{opadd}) the O(N) tensor-structure arising from (\ref{tensstr}). The
corresponding two possible $O(N)$ symmetric operators
\begin{eqnarray}
\tb\tb O_1=\left(\sum\limits_{i=1}^N\ph_i^2\right)\O(i\d)
\left(\sum\limits_{i=1}^N\ph_i^2\right),\nn
\tb\tb O_2=\sum\limits_{i,j=1}^N\ph_i\ph_j\O(i\d)\ph_i\ph_j
\end{eqnarray}
have the relative weight $N+4:4$, so for a general $\phi$-configuration
we have to add the following
operators to the basic action (\ref{p4}):
\begin{equation}
\Delta{\cal L}_{3D}={\lambda^2\over128\pi^2}\left({N+4\over9} O_1 +
{4\over9} O_2\right).
\label{delL}\end{equation}
This is the central proposition of this paper.

These operators are generated at one-loop level, so for two loop calculations
of the effective potential they must be treated at one loop in the three
dimensional theory. Whence the one loop contributions of different modes are
independent we can calculate the extra divergences produced by
 these operators from  the Lagrangian
\begin{equation}
{\cal L}_{3D}=\sum\limits_{i=1}^N\fel\left((\d\ph_i)^2+m_T^2\ph_i^2\right)
+\Delta{\cal L}_{3D}.
\end{equation}
Shifting the fields $\ph_i\to\ph_i\,(i\neq N)$ and $\ph_N\to\ph_N+\P$
respectively, and using the fact $\O(k=0)=0$, we get for the quadratic terms:
\begin{eqnarray}
\tb{\cal L}^{(2)}_{3D}=\tb\sum\limits_{i=1}^{N-1}\fel\left((\d\ph_i)^2+m_T^2
\ph_i^2+{\lambda^2\P^2\over72\pi^2}\ph_i\O\ph_i\right)\nn
\tb\tb+\fel\left((\d\ph_N)^2+m_T^2\ph_N^2+{(N+8)\lambda^2\P^2\over144\pi^2}
\ph_N\O\ph_N\right).
\end{eqnarray}
The effective potential using the above expression is:
\begin{eqnarray}
v_{eff}\tb=\tb\fel m_T^2\P^2+(N-1)\fel\int\limits_\Lambda{d^3k\over
(2\pi)^3}\ln\left[k^2+m_T^2+{\lambda^2\P^2\over72\pi^2}\O(k)\right]\nn
\tb\tb+\fel\int\limits_\Lambda{d^3k\over(2\pi)^3}\ln\left[k^2+m_T^2+
{(N+8)\lambda^2\P^2\over144\pi^2}\O(k)\right].
\end{eqnarray}
After expanding it with respect to $\P^2$ the contribution to the quadratic
term is:
\begin{equation}
v_{eff}^{(2)}=\fel m_T^2\P^2+\left({N+2\over3}\right){\lambda^2\P^2
\over64\pi^4}\int\limits_0^\Lambda\!dk\,{k^2\over k^2+m_T^2}\O(k).
\end{equation}
The leading divergencies are provided by
\begin{equation}
v_{eff}^{(2)div}=\left({N+2\over3}\right)
{\lambda^2\P^2\over64\pi^4}\int\limits_0^\Lambda\!dk\,
\left[\log\left({k\over T}\right)+{\pi^2Tk\over T^2+k^2}
+F\left({k\over\Lambda}\right)\right],
\end{equation}
what after performing the integrations simplifies to
\begin{equation}
v_{eff}^{(2)div}=\left({N+2\over3}\right)
{\lambda^2\P^2\over64\pi^4}\left[\Lambda\log\Lambda+\pi^2T\log\Lambda
-\Lambda\left(\ln4\pi+1-\gamma_E+\int\limits_0^1\!dx\,{2-x\over x}
\ln{2-x\over2}\right)\right].
\label{divnonloc}\end{equation}
The divergencies of the three dimensional model up to two loop level have to
be cancelled against the induced counterterm $\delta m^2_{ind}$ (see
(\ref{pred})). There are divergencies coming from the local terms
(\ref{divloc}) and others coming from the nonlocal terms
(\ref{divnonloc}). Indead, summing these terms all
divergencies, including the linear ones are cancelled exactly.
This means, that for the proper treatment
of the divergencies (or for the proper treatment of the scale dependence
of the reduced theory) we must include the new
operators (\ref{delL}) into the effective description .
These are nonlocal operators, the nonlocality is of
the order ${\cal O}(T^{-1})$ in the coordinate space. On scales larger than
${\cal O}(T^{-1})$ these can be omitted. So a coarse grained local action with
grain size ${\cal O}(T^{-1})$ may be a good choice to examine the whole theory
\cite{Mack} -- on this scale the theory forgets of its 4 dimensional origin.
In the course of coming down from scale $\Lambda$ to scale ${\cal O}(T)$ in
the momentum space the nonlocal operators modify the coupling constants of the
local ones. The actual magnitude of this modification can be estimated only
after doing this calculation, for instance, by matching some n-point functions.
This is, however, the task of future calculations.

\section{Conclusions}

In this paper we advocated a non-local effective 3D theory, equivalent to the
finite temperature O(N) symmetric scalar theory on the 2-loop level. The
proposed theory has the form
\begin{equation}
{\cal L}_{eff}={1\over 2}[(\partial_i\phi_\alpha )^2+m^2(T)\phi_\alpha ^2]
+{\lambda\over 24}(\phi_\alpha ^2)^2
+{\lambda^2\over 128\pi^2}[{N+4\over 9}\phi_\beta ^2\Omega (i\partial )
\phi_\gamma ^2+{4\over 9}\phi_\beta\phi_\gamma\Omega (i\partial )
\phi_\beta\phi_\gamma ],
\label{nonloc}
\end{equation}
with $m^2(T)$ given by the coefficient of $\Phi_0^2/2$ in (\ref{numred}),
and $\Omega (i\partial )$ defined in (\ref{omega}). Now with a further step
one can define a local approximation to this theory, when all static degrees
of freedom are integrated out with $k >> T$. A local cut-off theory with
$\Lambda_3=\kappa T, \kappa\sim 1-5$ will be arrived at what is very close in
spirit
to the effective theory approach of \cite{Braaten}. The simplest version
of this theory has approximately the form
\begin{equation}
{\cal L}_{eff,eff}={1\over 2}[Z(\partial_i\phi_\alpha )^2+(M^2(T)+\Sigma )
\phi^2_\alpha ]+{\lambda_3+\Delta \lambda_3\over 24}\phi^4,
\label{locth}
\end{equation}
where $Z,\Sigma ,\Delta\lambda_3$ should be found by matching some important
quantities calculated from the respective theories (\ref{nonloc}) and
(\ref{locth}). We propose to calculate the couplings of (\ref{locth})
by matching the low-$k$ behaviors of the 2-point functions and the
self-coupling (4-point function). The matching is achieved the easiest way
in the symmetric (high-T) regime, where $m^2(T)>0$.

In the non-local theory our construction ensures the finiteness of the
2-point and 4-point functions. In the cut-off theory the dependence on the
cut-off is not absorbed by anything, but is expected to be very weak
if $\kappa$ is varied around unity. It will be very interesting to see
the relation of the result of this approach to that of \cite{Braaten}.

The practical interest of this second approximation is its easy discretisation
into a lattice theory. However, since it is a cut-off theory, it should not
be understood as a continuum theory, but it is better to study it with finite
lattice constant of ${\cal O}(T^{-1})$ (coarse grained lattice). This means
that the thermodynamics of this model should be studied with a dimensionless
temperature $\Theta =aT$ varying around unity. This circumstance
seems to be very advantageous in view of the very weak phase transition
signals, experienced in the small $\Theta$ region in some effective models
\cite{Karsch}.

The analysis of the present paper should be carried over to gauged
Higgs theories. The vector nature of the gauge potential does not present
serious difficulty. More important is the task of analysing another class
of diagrams contributing to the non-local behaviour of the effective theory,
namely the four-leg box diagrams with non-vanishing external spatial momenta.

\bigskip\bigskip
\centerline{\bf ACKNOWLEDGEMENTS}
\bigskip
The author would like to express his special gratitude to Andr\'as Patk\'os
for his continous advice and help in writing this paper. He would
also like to thank Z.Fodor, K.Kajantie and M.Laine for valuable discussions
and comments.

\appendix

\section{Computing threefold cut-off integrals in 3 dimensions}

We shortly discuss the method of performing 3D integrals relevant to the
evaluation of the setting sun diagram contribution both in 3 and 4 dimensions.
Here we use the regularisation convention that momenta on any internal line
should not exceed the value of the cut-off, separately. (Another natural
procedure would be to cut-off the loop momenta.)
The integral to be performed is:
\begin{equation}
I=\int\!{d^3p\over(2\pi)^3}{d^3q\over(2\pi)^3}{d^3k\over(2\pi)^3}
\Theta(\Lambda-|p|)\Theta(\Lambda-|q|)\Theta(\Lambda-|k|)(2\pi)^3
\delta^{(3)}(p+q+k)\,f(p^2,q^2,k^2)
\end{equation}

Let us perform first the  $k$-integral with help of the delta-function -- then
instead of $k^2$ we can write $(p+q)^2$. The remaining integral depends on
$p^2$, $q^2$ and $x^2=(pq)^2/(p^2q^2)$ ($-1<x\leq1$); $k^2=p^2+q^2+2|p||q|x$.
One can replace x by the variable $k$ (its allowed range of variation is
$|p-q|<k<p+q$), then the integral will be symmetric in $p,\,q,\,k$, with the
domain of integration
\begin{equation}
\Delta:=\{k<p+q\qquad p<q+k\qquad q<k+p\}.
\end{equation}

$$\vbox{
\setlength{\unitlength}{0.015in}
\begin{picture}(0,100)(40,700)
\put(180,700){\line(1,0){100}}
\put(180,700){\line(0,1){100}}
\multiput(280,700)(-0.5,-0.5){7}{\makebox(0.4444,0.6667){\sevrm .}}
\multiput(280,700)(-0.5,0.5){7}{\makebox(0.4444,0.6667){\sevrm .}}
\multiput(180,800)(-0.5,-0.5){7}{\makebox(0.4444,0.6667){\sevrm .}}
\multiput(180,800)(0.5,-0.5){7}{\makebox(0.4444,0.6667){\sevrm .}}
\put(270,703){\line(0,-1){6}}
\put(183,790){\line(-1,0){6}}
\put(270,692){\makebox(0,0)[t]{\raisebox{0pt}[0pt][0pt]
{$\scriptstyle\Lambda$}}}
\put(174,786){\makebox(0,0)[rb]{\raisebox{0pt}[0pt][0pt]
{$\scriptstyle\Lambda$}}}
\multiput(270,700)(0,2){30}{\makebox(0.4444,0.6667){\sevrm .}}
\multiput(180,790)(2,0){30}{\makebox(0.4444,0.6667){\sevrm .}}
\put(210,703){\line(0,-1){6}}
\put(183,730){\line(-1,0){6}}
\put(210,692){\makebox(0,0)[t]{\raisebox{0pt}[0pt][0pt]
{$\scriptstyle p$}}}
\put(174,728){\makebox(0,0)[rb]{\raisebox{0pt}[0pt][0pt]
{$\scriptstyle p$}}}
\put(179,805){\makebox(0,0)[rb]{\raisebox{0pt}[0pt][0pt]
{$q$}}}
\put(285,697){\makebox(0,0)[lb]{\raisebox{0pt}[0pt][0pt]
{$k$}}}
\thicklines
\put(210,700){\line(-1,1){30}}
\put(210,700){\line(1,1){60}}
\put(180,730){\line(1,1){60}}
\put(270,760){\line(0,1){30}}
\put(240,790){\line(1,0){30}}
\end{picture}
}$$
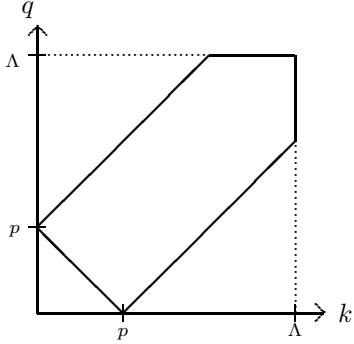
\centerline{{\bf Fig.2} \quad {\sl Integration range at fixed p}}
\bigskip

Our expression after the angular integrations is of the form
\begin{equation}
I={1\over8\pi^4}\int\limits_\Delta\!dpdqdk\,\Theta(\Lambda-p)
\Theta(\Lambda-q)\Theta(\Lambda-k)\,pqk\,f(p^2,q^2,k^2).
\end{equation}
or writing out the limits of integration (cf. Fig.2)
\begin{equation}
I={1\over8\pi^4}\left[\int\limits_0^\Lambda\!dp\int\limits_0^\Lambda\!dq
\int\limits_{|p-q|}^{p+q}\!dk-\int\limits_0^\Lambda\!dp
\int\limits_{\Lambda-p}^\Lambda\!dq\int\limits_\Lambda^{p+q}\!dk\right]
pqk\,f(p^2,q^2,k^2)
\end{equation}
Often happens, that the integrand ($f$) is symmetric to the interchange of
$p$ and $q$. In this case we can write a simpler formula:
\begin{equation}
I={1\over4\pi^4}\left[\int\limits_0^\Lambda\!dp\int\limits_0^p\!dq
\int\limits_{p-q}^{p+q}\!dk-\int\limits_{\Lambda/2}^\Lambda\!dp
\int\limits_{\Lambda-p}^p\!dq\int\limits_\Lambda^{p+q}\!dk\right]
pqk\,f(p^2,q^2,k^2)
\end{equation}

\section{Computation of the setting sun in 4 dimensions}

The integral we have to evaluate is (\ref{eq:setsun})
\begin{equation}
K(m_p,m_q,m_k)=\int\!\!{d^3pd^3qd^3k\over(2\pi)^9}T^3\sum\limits_{n,m,l\neq0}
\!\!{\beta\delta_{n+m+l}(2\pi)^3\delta(p+q+k)\over(\omega_n^2+E_p^2)
(\omega_m^2+E_q^2)(\omega_l^2+E_k^2)},
\end{equation}
where $E_p^2=p^2+m_p^2$. The summations can be performed with help of
the ''Saclay-method'' \cite{Pisarski}. We get the expression:
\begin{equation}
K(m_p,m_q,m_k)=\int\!\!{d^3pd^3qd^3k\over(2\pi)^6}\delta(p+q+k)D(E_p,E_q,E_k),
\label{eq:sacmet}\end{equation}
where
\begin{eqnarray}
D(E_p,E_q,E_k)\tb=\tb{2T^2\over E_p^2E_q^2E_k^2}+{1\over4E_pE_qE_k
(E_p+E_q+E_k)}\nn
\tb\tb-{T\over E_pE_qE_k}\left[D_1(E_p,E_q,E_k)+D_1(E_q,E_p,E_k)
+D_1(E_k,E_q,E_p)\right]\nn
\tb\tb+{1\over4E_pE_qE_k}\left[
D_2(E_p,E_q,E_k)+D_2(E_q,E_p,E_k)+D_2(E_k,E_q,E_p)\right]\nn
\tb\tb+{1\over4E_pE_qE_k}\left[
D_3(E_p,E_q,E_k)+D_3(E_q,E_p,E_k)+D_3(E_k,E_q,E_p)\right],\nn
\label{eq:setsunparts}\end{eqnarray}
and
\begin{eqnarray}
D_1(E_p,E_q,E_k)\tb=\tb{n_{E_q}\over2E_p}\left({1\over E_q+E_k}-
{1\over E_q-E_k}\right)+(q\leftrightarrow k)+{1\over 2E_p(E_q+E_k)},\nn
D_2(E_p,E_q,E_k)\tb=\tb n_{E_p}\left({1\over E_p+E_q+E_k}-
{1\over E_p-E_q-E_k}\right),\nn
D_3(E_p,E_q,E_k)\tb=\tb n_{E_q}n_{E_k}\biggl({1\over E_p+E_q+E_k}
+{1\over E_p-E_q+E_k}+{1\over E_p+E_q-E_k}\nn
\tb\tb\qquad\qquad+{1\over E_p-E_q-E_k},\biggr)
\end{eqnarray}
and $n_E$ is the Bose-Einsein function ($n_E=(e^{\beta E}-1)^{-1}$).
The parts depending on T only in the form $n_E$ agree with the expression
of Parwani \cite{Parwani}.

In the course of the integration of each single term in (\ref{eq:setsunparts})
UV and various IR divergence problems arise. In order to control them the best
is
 to work from the start with a regularized propagator in the mixed $(\tau,p)$
representation \cite{Pisarski}:
\begin{equation}
\Delta_{reg}(\tau,E_p)=\Delta(\tau,E_p+\ep)\Theta(\Lambda-|p|),
\end{equation}
where
\begin{equation}
\Delta(\tau,X)={n_X\over2X}\left(e^{(\beta-\tau)X}+e^{\tau X}\right).
\end{equation}
With this propagator, using the general scheme to perform threefold
integrals in 3 dimensions (Appendix A) we get the following expression
for (\ref{eq:sacmet}):
\begin{equation}
K_{reg}={1\over8\pi^4}\int\limits_{\Delta}\!\!dpdqdk\,pqk\,
\Theta(\Lambda-p)\Theta(\Lambda-q)\Theta(\Lambda-k)
D(E_p+\ep,E_q+\ep,E_k+\ep).
\end{equation}
The full expression is IR safe, since we omitted all static parts.
Therefore at the end of the calculations we may perform the $\ep\to0$
limit.

If we rescale all the variables by T, the full integral $K\sim T^2$. At
high temperatures it is possible and convenient to expand the integral
with respect to the masses:
\begin{equation}
K_{reg}(m_p,m_q,m_k)=T^2K_{reg}(0,0,0)+m_p^2{\d\over\d m_p^2}
K_{reg}(m_p,m_q,m_k)\biggr|_{m_i=0}\!\!\!\!\!+(p\leftrightarrow q)
+(p\leftrightarrow k)+\dots
\label{Kreg0}\end{equation}
The omitted parts are proportional to $m^4/T^2$.

The regularization of the propagator regularizes all the arising
divergencies, which makes possible the correct derivation of the
finite parts of the integrals. This is important in the case of
$K_{reg}(0,0,0)$, where the finite part is multiplied by $T^2$, which means,
that it cannot be absorbed by counterterms in any T-independent
renormalization scheme.

For the calculation of $K_{reg}(0,0,0)$ I followed the following
general ideas. First one has to separate the (UV/IR) divergent parts.
They should have the simplest possible form allowing anlytic evaluation
of the coefficients of the divergencies.  Substracting the divergent
pieces we can perform the  $\Lambda\to\infty$ and the $\ep\to0$ limits
respectively. If there is some Bose-Einsein functions ($n_E$) involved
in the integral containing $\Lambda$ or $\ep$, we can use the relations:
\begin{eqnarray}
\tb\tb\lim\limits_{\ep\to0}\int\!dx\,\ep n(\ep x)\,f(x)=\int\!dx\,
{f(x)\over x}\qquad{\rm if\,it\,exists,}\nn
\tb\tb\lim\limits_{\Lambda\to\infty}\int\!dx\,\Lambda^2n_{\Lambda x}\,f(x)
={\pi^2\over6}f'(0)\qquad{\rm if}\,f(0)=0.
\end{eqnarray}
After this step exclusively integrals over rational functions remain to be
performed. The $k$-integration can be done analytically. The $p$,$q$
integrations have been transformed to a form (in general by
introducing new integrational variables), where at most 1 numerical
integration was left.

Armed with these ideas all the integrals in $K_{reg}(0,0,0)$ can be
performed. Let us denote the integrational measure in (\ref{Kreg0})
after shifting all the variables by $\ep$ by
\begin{equation}
\int\!dM_{\ep}={1\over8\pi^4}\left[\int\limits_\ep^\Lambda\!dp(p-\ep)\!
\int\limits_\ep^\Lambda\!dq(q-\ep)\!\!\!\!\!\!\int\limits_{|p-q|+\ep}
^{p+q-\ep}\!\!\!\!\!\!dk(k-\ep)\,\,-\,\,\int\limits_0^\Lambda\!dpp\!\!
\int\limits_{\Lambda-p}^\Lambda\!\!dqq\!\!\int\limits_\Lambda^{p+q}\!\!dkk\,
\right]
\end{equation}
In the second part of the measure no IR divergencies occure, the
$\ep\to0$ limit can be taken directly. The results of the relevant
integrals are
(ignoring terms, which vanish as $\Lambda\to\infty$ or $\ep\to0$):
\begin{eqnarray}
\tb\tb I_A=\int dM_{\ep}{1\over p^2q^2k^2}={1\over16\pi^2}
\ln{\Lambda\over\ep}-0.0190158229,\nn
\tb\tb I_B=\int dM_{\ep}{1\over 2pqk^2(p+q)}=0.0009950146\,\Lambda,\nn
\tb\tb I_C=\int dM_{\ep}{1\over 2pqk^2}\left({n_p+n_q\over p+q}
+{n_q-n_p\over p-q}\right)={1\over16\pi^2}\ln{1\over\ep}-0.0115413434,\nn
\tb\tb I_D=\int dM_{\ep}{1\over 4pqk(p+q+k)}=0.0001041333\,\Lambda^2,\nn
\tb\tb I_E=\int dM_{\ep}{n_p\over 4pqk(p+q+k)}={1\over192\pi^2}\ln{\Lambda}
+0.0000776673,\nn
\tb\tb I_F=\int dM_{\ep}{n_p\over 4pqk(q+k-p)}={1\over192\pi^2}
\ln{\Lambda\over\ep}+0.0003657838,\nn
\tb\tb I_G=\int dM_{\ep}{n_pn_q\over 4pqk}\left({1\over k+p+q}
+{1\over k+p-q}+{1\over k-p+q}+{1\over k-p-q}\right)=\nn
\tb\tb\qquad\qquad={1\over64\pi^2}\ln{1\over\ep}-0.0044038354.
\end{eqnarray}
With these integrals we can express we can express the first term in
the expansion of (\ref{Kreg0}):
\begin{equation}
K_{reg}(0,0,0)=2I_A-3I_B-3I_C+I_D+3I_E+3I_F+3I_G
\end{equation}
The IR divergencies cancel each other as we expected, the rest gives the
result appearing in the expression (\ref{constants}):
\begin{equation}
K_{reg}(0,0,0)=0.0001041333\,\Lambda^2-0.0029850437\,\Lambda T+
{5\over32\pi^2}T^2\loT{\Lambda}-0.0152887686\,T^2.
\end{equation}

Now, it is worthwhile to give a few commnets on
the second term  of the mass-expansion. Its finite
part can be absorbed into the finite renormalization, so only the divergent
pieces are
important. Therefore less terms are to be preserved.
The differentiation with respect to the mass squared can be changed
to differentiation with respect to the energy or (since we put afterwards
 all masses $=0$) to the momentum under the integral:
\begin{equation}
{\d\over\d m_k^2}\left(\int_{\Lambda,\ep}\!\!D\right)\biggr|_{m_i=0}=
\int_{\Lambda,\ep}{1\over2E_k}{\d D\over\d E_k}\biggr|_{m_i=0}
\end{equation}
When the derivation acts on $n(E)$, we can integrate partially, so at the end
only rational functions and eventually $n(E)$ factors remain, and with the
previous ideas the integrations can be reduced to one-dimensional
 numerically computable integrals.

\end{document}